\documentclass[epj]{webofc}
\usepackage[varg]{txfonts}

\usepackage{amsmath,amssymb}
\usepackage{graphicx}
\usepackage{epsfig}
\usepackage{slashed}
\usepackage{color}

\woctitle{QUARKS}

\newcommand{\bq}{{\bf q}}
\newcommand{\be}{\begin{equation}}\newcommand{\ee}{\end{equation}}
\newcommand{\bea}{\begin{eqnarray}}\newcommand{\eea}{\end{eqnarray}}

\newcommand{\om}{\omega}

\newcommand{\G}{\Gamma}

\renewcommand{\phi}{\varphi}
\def\Perp{{\scriptscriptstyle \| }}
\newcommand{\p}{p_{\Perp}}

\def\bp{\mathbf{p}_\Perp}
\def\bq{\mathbf{q}_\Perp}
\def\({\left(}
\def\){\right)}
\def\[{\left[}
\def\]{\right]}
\def\tr{{\mathop{\rm tr}}}
\def\={\mathop{=}}
\def\seq{\mathop{\simeq}}
\renewcommand\Re{\mathop{\rm Re}}
\newcommand{\xx}{{\rm xx}}
\newcommand{\Ref}[1]{(\ref{#1})}

\def\a{\alpha}

\def\g{\gamma}
\def\tk{\varkappa}
\def\S{\sigma_{22}}

\begin{document}

\title{
%On applications of Quantum Field Theory in physics of graphene\\
%Applying Quantum Field Theory to graphene\\
%Quantum Field Theory as a looking-glass for graphene\\
Graphene through the looking glass of QFT%
\footnote{Based on a talk given by
D.~V.~Vassilevich at QUARKS16, Pushkin 2016.}}

\author{\firstname{Ignat V.} 
        \lastname{Fialkovsky}\inst{1}
	  \fnsep\thanks{\email{ifialk@gmail.com}}
	  \and
        \firstname{Dmitri V.} 
	\lastname{Vassilevich}\inst{1}
	  \fnsep\thanks{\email{dvassil@gmail.com}} 
        % etc.
}
\institute{
CMCC, Universidade Federal do ABC, Santo Andr\'e, S.P.,
Brazil
          }

\abstract{
This paper is aimed to review and promote the main applications of the methods of Quantum Field Theory to description of quantum effects in graphene. We formulate the effective electromagnetic action following from the Dirac model for the quasiparticles in graphene  and apply it for derivation of different observable effects like the induced mean charge, quantized conductivity, Faraday effect, and Casimir interaction involving graphene samples.
%\keywords{Graphene; Dirac model.}
}

%\ccode{PACS numbers: 73.63.-b, 11.10.Kk}

\maketitle

%%%%%%
\section{Introduction}
Since its discovery some 12 years ago \cite{Novoselov2004} graphene gained worldwide recognition and contributed both to our futuristic dreams of nanoscale computing and rather earthly to development of new theoretical methods in condensed matter physics. On the other hand, due to its very peculiar property of having pseudo-relativistic quasi-particle excitations, graphene provided an all new playground for application of the methods and approaches of Quantum Field Theory (QFT).

In \cite{FV2011} we already presented an overview of the main achievements of field-theoretical methods in graphene physics at that moment, such as explanation of anomalous Hall Effect in graphene, the universal optical absorption rate and the Faraday effect. It is the purpose of this paper to show the latest developments in the area and new applications of QFT in graphene physics.

The basis for all such considerations is the Dirac model for quasi-particles in graphene which was elaborated in full around 1984\cite{DIVM,Semenoff:1984dq}, some twenty years before actual discovery of graphene. Actually, all the basic properties of the model, such as linearity of the spectrum, were well known and widely used much earlier due to the 1947 paper by Wallace\cite{Wallace} who considered mono-atomic layers of carbon as constituents for describing band properties of graphite.
We redirect the interested reader to our short review \cite{FV2011}, or to more detailed ones, e.g. \cite{CastroNeto:2009zz}, for a complete derivation of the Dirac model, we present here only the main steps. 

\begin{figure}
\centering \includegraphics[width=6cm]{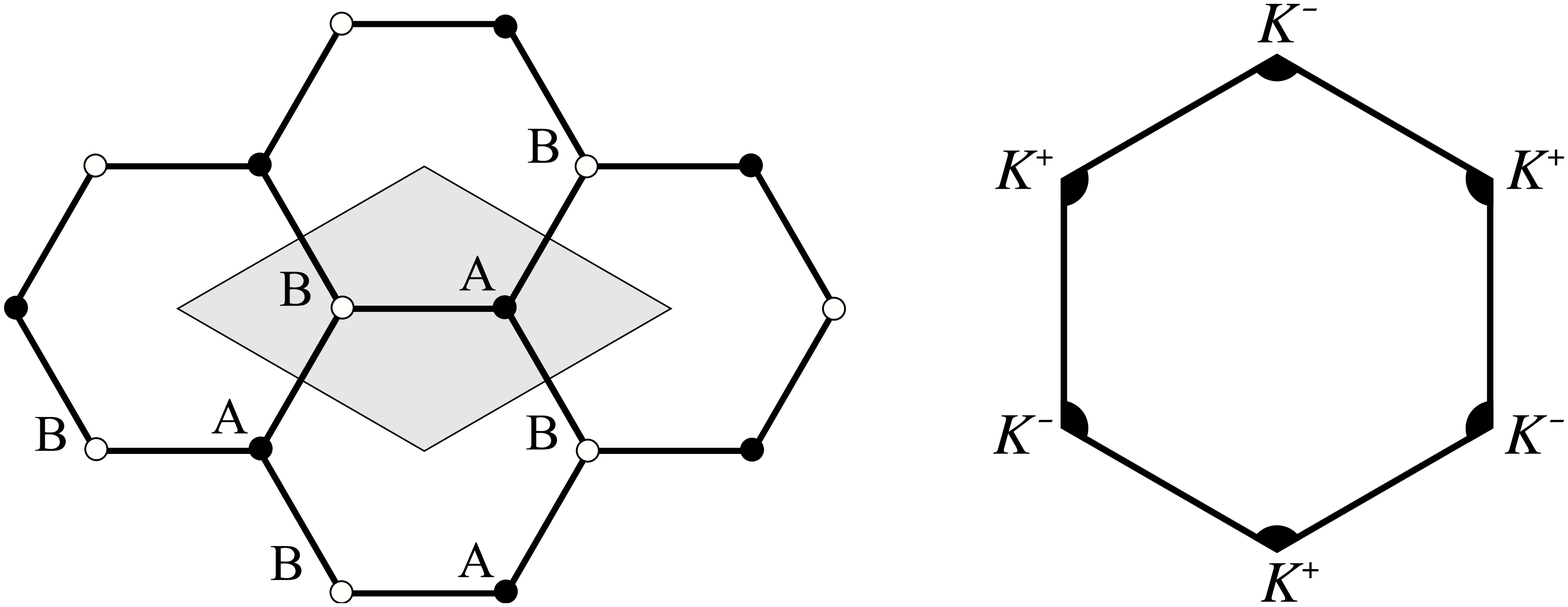} \caption{The honeycomb lattice
of graphene (left), and its first Brillouin zone (right).} \label{lattice}
\end{figure}

In graphene, the carbon atoms form a honeycomb lattice (see Fig.\ \ref{lattice}, left)
with two triangular sublattices A and B. The lattice spacing is $d=1.42$\AA.
The nearest neighbors of an atom from the sublattice A belong to the sublattice
B, and vice versa. In the \emph{tight binding model} only the interactions between electrons belonging to these nearest neighbors are taken into account. Considering a very simple process of hoping of the weekly bonded $\pi$-electrons between the nearest neighbors, in the linear approximation in momenta one obtains the Hamiltonians
\begin{equation}
 H_\pm = \frac{3td}2 \left( \begin{array}{cc} 0 & iq_1\pm q_2 \\
-iq_1\pm q_2 & 0    \end{array} \right) =
v_F (-\sigma_2 q_1 \pm \sigma_1 q_2)\,,\label{Hpm}
\end{equation}
acting on the electron envelope function $\psi_\pm = (\psi_A^\pm,\psi_B^\pm)$.
Here $t$ stands for hopping parameter, $v_F=(3td)/2$ is the Fermi velocity, $\sigma_i$ are the standard Pauli matrices. With the lower index $\pm$ we distinguish two inequivalent corners $K^\pm$, called Dirac points, of the first Brillouin zone corresponding to the crystal lattice of graphene.

One can introduce now a four-component spinor $\psi\equiv (\psi_A^+,\psi_B^+,\psi_A^-,\psi_A^-)$, to build a single Dirac Hamiltonian from the $H_\pm$ blocks
\be
  H=-i v_F\gamma^0 \gamma^a \partial_a,\quad a=1,2,
  \label{H}
\ee
where the momenta $i\bq$ is replaced by partial derivatives, and
\begin{equation}
\gamma^0=\left( \begin{array}{cc} \sigma_3 & 0 \\ 0 & \sigma_3 \end{array}
\right),\qquad
\gamma^1=\left( \begin{array}{cc} i\sigma_1& 0 \\ 0 & i\sigma_1 \end{array}
\right),\qquad
\gamma^2=\left( \begin{array}{cc} i\sigma_2& 0 \\ 0 & -i\sigma_2 \end{array}
\right).\label{gam}
\end{equation}
These $4\times 4$ gamma matrices are taken in a reducible representation
which is a direct sum of two inequivalent $2\times 2$ representations of the Clifford algebra in $2+1$ dimensional space. 

The wave-functions of graphene quasiparticles in each of these two-component representations are similar to the 2-spinors describing electrons in QED$_{3+1}$. However, in the case of graphene this {\it pseudo}\,spin index refers to the sublattice degree of freedom rather than the real spin of the electrons. The latter did not appear up to now in presented Dirac model, it results just in doubling the number of spinor components, so that the complete wave functions of quasiparticles in graphene are given by 8-component spinors  ($N=4$ species of two--component fermions).

The tight binding model takes into account just the simplest interaction that appears in graphene. It can thus be improved by including other couplings as the next-nearest neighbor coupling \cite{Gusynin:2007ix}, for instance.  All in all, it is expected that the Dirac model, which approximates the tight binding model at ``low energies'' is valid under suitable modifications at least until the energies of $\sim 2\, {\rm eV}$.

Up to now, we were only considering pristine freestanding graphene, which of course is not the case in may practical situations. The quasi-particles can have a non-zero mass $m$ (due to the interaction with the substrate or other effects\cite{Appelquist,massgap1}), and be considered away from the charge neutrality point, i.e. in presence of the non-vanishing chemical potential $\mu$. While the generation of mass is an open controversial question, the generation of $\mu$ is a straightforward procedure either by applying gate potential to graphene samples, depositing them on a substrate or by molecular doping \cite{coletti10,riedl10}. Moreover, most of real experiments are performed at non-zero temperature, and none of graphene samples can be considered completely free of impurities. While the temperature of a sample is easily taken into account by usual rules of real or imaginary time thermal field theory, introduction of impurities is a much more complicated issue\cite{mirlin06}. In the simplest cases (relevant to applications of QFT methods) they can be described by a phenomenological parameter $\Gamma$, see Eq. (\ref{subs}) below.

%%%%%%%%
\section{Effective electromagnetic action}\label{sec-action}
It is evident, that without interacting with electromagnetic field, the properties of the (charged) quasi-particles in graphene would remain obscured for any experimentalist. To introduce such interaction (which in many quantum mechanical calculations is understood but not explicified) one should replace in \Ref{H} the usual partial derivatives by gauge-covariant ones: $\partial\to\partial+ieA$. Note that the electromagnetic potential is not confined to the graphene surface, but rather propagates in
the ambient $3+1$ dimensional space.  

The full action of the system derived in this way with graphene assumed laying at $x^ 3=0$ reads
\be
	S= -\frac{1}4 \int d^4x\, F^2_{\mu\nu}+\int d^4x\, \delta(x^3)\,\bar\psi \slashed{D}\psi 
\label{Sfull}
\ee
here $F_{\mu\nu}=\partial_\mu A_\nu-\partial_\nu A_\mu$,  $\mu,\nu =0,1,2,3$ and
\begin{equation}
\slashed{D}=i\tilde\gamma^m (\partial_m +ieA_m)+\dots \label{SDsl}, \qquad m=0,1,2,
\end{equation}
and the dots denote any of additional terms described at the end of the previous
section. Tilde over $\gamma$ means rescaling of the space-components 
\begin{equation}
\tilde\gamma^0=\gamma^0,\qquad \tilde\gamma^{1,2}=v_F\gamma^{1,2} \,.\label{tilded}
\end{equation}
In natural units $\hbar=c=1$ the Fermi velocity $v_F=1/300$ thus bringing the second small parameter in the theory alongside with the fine structure constant $\alpha= e^2/(4\pi) \simeq 1/137$.

By functional integration over the quantized fermions in the Dirac sector of the model \Ref{Sfull}, an effective action for electromagnetic field $A_\mu$ can be obtained
\be
S= -\frac{1}4 \int d^4x\, F^2_{\mu\nu}+S_{\!\rm eff}(A),\qquad 
		S_{\!\rm eff}(A)\equiv-i\ln \det \slashed{D}\,.
\label{Seff}
\ee
Here, $det$ stays for a functional determinant of a differential operator. $S_{\!\rm eff}(A)$ is a sum of one-loop Feynman diagrams with an arbitrary number of external photons. This action, or rather its expansion in powers of $A$ gives rise to the description of electronic and optical properties of graphene and graphene based devices.

In particular, the mean density of fermionic current, $\langle j^m\rangle=\langle\bar\psi\tilde\g^m\psi\rangle$, is defined as the first functional derivative of $S_{\rm eff}(A)$ 
\be
\langle j^m (x)\rangle=-\left[ \frac \delta{\delta A_m(x) }\, S_{\rm eff}(A)\right]_{A=0}
=  -i \, {\rm tr} (\tilde\gamma^m \mathcal{S}(x,x))\,.\label{j01}
\ee
The electronic conductivity tensor of graphene devices can be obtained via the polarization tensor, given by the second functional derivative of  $S_{\!\rm eff}(A)$
\be
\Pi^{mn} (x,y) =\left[ \frac{\delta^2}{\delta A_m(x) \delta A_n(y) }\, S_{\rm eff}(A)\right]_{A=0}
=
    i  e^2  \tr\(\mathcal{S}(x,y) \tilde\gamma^m \mathcal{S}(y,x) \tilde\gamma^n\),
    \quad m,n=0,1,2.
  \label{Pi_g}
\ee
In \Ref{j01} and \Ref{Pi_g} the notation $\mathcal{S}(x,y)$ stands for the free fermion propagator
\begin{equation}
\mathcal{S}(x,y)\equiv \slashed{D}^{-1}(x,y) \big|_{A=0}.
    \label{S}
\end{equation}
Depending on the nature of particular graphene based system under consideration its  calculation can represent a separate very interesting physical problem,
e.g. \cite{Nalimov}.

A very similar effective electromagnetic action was also derived for topological insulators \cite{QiShang2011}. It is based on the fact that the conducting surface modes of such materials are also described by Dirac model \cite{IT}.

\subsection{Fermion Propagator}
For the pristine infinite (and thus translation invariant) graphene the propagator is expressed in Fourier transform as
\be
\mathcal{S}(x,y) \equiv \mathcal{S}(x-y)
   =i\int d^3 p\, e^{i p(x-y)}  \mathcal{S}(p)
\ee
where $\mathcal{S}(p)$ has all familiar form
\be
  \mathcal{S}(p) = \frac{1}{\tilde\gamma^ j p_ j+v_F m}.
  \label{S0}
\ee
We include the mass gap $m$ both for the completeness of our description and to be able to use further on the Pauli-Villars UV regularization. 

The simplest modifications which expression \Ref{S0} permits are concerned with introducing the chemical potential $\mu$ (i.e. considering the graphene away from the charge neutrality point) and weak short range impurities characterized by phenomenological parameter $\G >0$. Both this alterations are achieved by the following substitution in \Ref{S0}
\begin{equation}
    p_0 \to \zeta(p_0)\equiv p_0 +\mu+ i \Gamma{\rm\ sgn}p_0.
    \label{subs}
\end{equation}

In more elaborate systems, for example for nanoribbons where translation invariance is reduced in one of the directions, say $x^1$, the fermion propagator can be obtained as a sum over the eigenmodes $\psi $ of ${\cal D}=\gamma^0\slashed{D} |_{A=0}$
\be
  {\cal D}\psi = \mathcal{E}\psi
  \label{EVP}
\ee
in the following form:
\begin{equation}
\mathcal{S}(x,y)=
    \sum_{n}\int dp_0\, dp_2\,
    \frac{\psi_{n,p_0,p_2}(x)\otimes\bar\psi_{n,p_0,p_2}(y)}
    {\mathcal{E}(n,p_0,p_2)},
    \label{S}
\end{equation}
here $n$ stands for a discreet (multi)index describing the fermion degrees of freedom perpendicular to the ribbon, while $p_0$ and $p_2$ are familiar continuous momenta components.

The Dirac operator $\slashed{D} |_{A=0}$ in \Ref{EVP} should be supplied with appropriate boundary conditions. One of the possibilities is the so called Berry-Mondragon boundary conditions \cite{BerryM} which for a nanoribbon of width $W$ read
\begin{eqnarray}
%  to remove numbering (before each equation)
  \psi=-i\g^1 \psi  && {\rm at\ } x^1=0 \nonumber\\%
  \psi=+i\g^1 \psi  && {\rm at\ } x^1=W \label{BeM}.
\end{eqnarray}
The discrete index $n$ is given now by \cite{BFSV2014}
\be
    n\equiv\{\alpha, p_1\}\quad 
    \alpha=\pm1, p_1=\frac{\pi}{2W} , \frac{3\pi}{2W},\ldots
      %\left(k-\frac 12 \right), k=1,2,\dots
    \label{k_1_m}
\ee
and the normalized spinor eigenmodes are
\begin{equation}
\psi_{\alpha,p_0,p_1,p_2}(x)=
     \frac{e^{-ip_0 x^0+ip_2x^2}}{2\pi \(\varkappa(\varkappa+\a p_2)\(W+\frac{m}{m^2+p_1^2}\)\)^{1/2}}
        \( \begin{array}{c}
            \a(\varkappa+\a p_2) \sin (p_1x^1) \\
            p_1 \cos (p_1x^1)+m\sin(p_1x^1)
            \end{array}
        \)\,, \label{modes}
\end{equation}
where $\varkappa=\sqrt{p_1^2+p_2^2+m^2}$.
 The eigenvalues $\mathcal{E}$ are
\begin{equation}
%    \mathcal{E}(p_0,p_2)=p_0 - v_F |p_2|,\qquad
        \mathcal{E}(\alpha,p_0,p_1,p_2)=p_0 +\alpha v_F \varkappa\,.
    \label{E}
\end{equation}
Chemical potential and weak impurities can be introduced into \Ref{S} using the same substitution \Ref{subs} in \Ref{E}. We will use \Ref{S} in the Sec. \ref{sec-ch} for calculation of the mean charge of Berry-Mondragon nanoribbons.

\subsection{Polarization operator and conductivities}
\label{sec-pol}
Particular importance of $\Pi^{mn}$ is due to the fact that this tensor
defines the electric current along graphene and the transmission of electromagnetic field which propagates through it. Indeed, the equations of motion following from the action \Ref{Seff} contain a singular term, localized on the graphene surface
\begin{equation}
\partial_\mu F^{\mu\nu} +\delta(x^3) \Pi^{\nu\rho}A_\rho =0.\label{Meq}
\end{equation}
We extended $\Pi^{mn}$ to a $4\times 4$ matrix with $\Pi^{3\mu}=\Pi^{\mu 3}=0$.

The first and most important observation is that the polarization tensor can be interpreted in terms of the in-plane conductivity of graphene. By its definition the conductivity is a coefficient (or a matrix in most general case) between the electric current $j$ and electric field $E$. The latter is related 
to the vector potential by $E_a =i\omega A_a$, $a=1,2$ in the
temporal gauge, $A_0=0$, and  with $\omega$ standing for the frequency.
In this way, one arrives at the field theoretical analog of the Kubo formula in Quantum Mechanics connecting the in-plane conductivities of a graphene layer and polarization operator \Ref{Pi_g}
\begin{equation}
\sigma_{ab}=\frac{\Pi_{ab}}{i\omega}\,, \quad a,b=1,2.\label{siPi}
\end{equation}
%Basing of this relation the 

Given the conductivity, $\sigma_{ab}$, or using \Ref{Meq} directly we are also able to investigate the optical properties of graphene by considering the scattering of the  electromagnetic field on the graphene samples. Indeed, (\ref{Meq}) describe a free propagation of electromagnetic field 
outside the surface $x^3=0$ subject to the matching conditions
\begin{eqnarray}
&&A_\mu \vert_{x^3=+0}=A_\mu\vert_{x^3=-0},\nonumber\\
&&(\partial_3A_\mu)\vert_{x^3=+0}-
(\partial_3A_\mu)\vert_{x^3=-0}=\Pi_\mu^{\ \nu}A_\nu \vert_{x^3=0}
\label{match}
\end{eqnarray}
on that surface. By solving these conditions (either in terms of electromagnetic potential or the fields themselves) for a given scattering problem we can obtain the reflection and transmission coefficients.
In our approach we obtained \Ref{match} directly from the effective electromagnetic action \Ref{Seff} but they can also be deduced from classical electrodynamics for a general conducting surface taking into account the relations \Ref{siPi}.

The polarization operator in $2+1$ dimensional systems has been investigated since at least 1980's \cite{Appelquist}. For a single
massive two-component fermion at zero temperature, zero chemical potential,
and without external electromagnetic fields simple calculations give
\begin{equation}
    \Pi^{mn}
        = \frac \alpha{v_F^2} \, \eta^{m}_{j}\left[
            \Phi(p) \left(g^{jl}-\frac{\tilde p^j\tilde p^l}{\tilde p^2}\right)
            + i \phi(p) \epsilon^{jkl}\tilde p_k
            \right]\eta_l^n \label{Pmn}
\end{equation}
where $\epsilon^{jkl}$ is the Levi-Civita totally antisymmetric tensor,
to $\epsilon^{012}=1$, $\eta_j^n={\rm diag}(1,v_F,v_F)$,
$\tilde p^m\equiv \eta_n^m p^n$. 
The tensor structure of $\Pi^{mn}$ is fixed by the symmetries, so that $\Pi^{mn}$ depends on two functions,
$\Phi$ and $\phi$, which read 
\begin{eqnarray}
&&\Phi(p)=
        \frac{2 m \tilde p -(\tilde p^2+4m^2){\rm arctanh }
({\tilde p}/{2m})}{2\tilde p},
    \label{Psi}\\
&&
\phi(p)
    =\frac{2m\, {\rm arctanh}(\tilde p/2 m )}{\tilde p}-1,\label{phi}
\end{eqnarray}
$\tilde p\equiv+\sqrt{\tilde p_j \tilde p^j}$, and $m>0$.
The origin of the pseudotensor part is very peculiar being directly linked to the parity anomaly, \cite{Dunne}. However, in real graphene without external fields the parity odd part of \Ref{Pmn} is canceled between different species of fermions, see \cite{Fialkovsky:2009wm}.  The cancellation occurs due to the form of gamma-matrices (\ref{gam}), containing two inequivalent representations related by the parity transformation. 
However, in parity odd systems, for instance in presence of external magnetic field or in-plane strain, the pseudo-scalar part of polarization tensor is non-trivial. The same is true for topological insulators \cite{IT}. 
 
In the presence of P-even matter, described by a non-zero chemical potential at non-zero temperature, the quasi-Lorentz invariance is broken, and the polarization tensor has a more complicated form than Eq.\ (\ref{Pmn}). All components of the polarization tensor can be expressed in this cases via two scalar form factors, for instance, $\Pi_{\rm tr}\equiv \Pi_{00}-\Pi_{11}-\Pi_{22}$ and $\Pi_{00}$ \cite{Zeitlin,Fialkovsky:2011pu}.
As in the QED/QCD cases,  these quantities consist of the vacuum part and a part carrying the dependence on $T$ and $\mu$. In the most general case derived up to now, with chemical potential, mass and non-zero temperature,  they are given by \cite{BFV2015}
\be \Pi_\xx(p;\mu, T) 	= \Pi_\xx^{(vac)}(p) 		+\Delta \Pi_\xx (p;\mu,T), 		\label{decomp1} \ee
where $\xx$ stands either for '${\rm tr}$' or '$00$', and the momentum is considered in Euclidean space, $p=(2 \pi n T, \bp)$, $n=1,2,\ldots$. The vacuum part, $\Pi_\xx^{(vac)}(p)$, corresponds to $\mu=T=0$. The matter part is given by
\be
\Delta\Pi_\xx =
\frac{ 8\alpha }{v_F^2}
	\int_m^\infty d\kappa
	\(1+ \Re\frac{M_\xx }{\sqrt{Q^2-4\p^2 (\kappa^2-m^2)}} \) \Xi (\kappa) .
\label{Delta Pi}
\ee
Here the distribution function, $\Xi\equiv {(e^{(\kappa+\mu)/T}+1)^{-1}} + {(e^{(\kappa-\mu)/T}+1)^{-1}}$, carries the dependence on $T$ and $\mu$. Further notations in \Ref{Delta Pi} are  
\begin{eqnarray} 
&&M_{00}	= -\tilde p^2+4i p_4 \kappa+4\kappa^2 , \nonumber\\ 
&& M_{\rm tr}=  -(2-v_F^2)(4m^2-\tilde p^2) +4(1-v_F^2)(p_4\kappa+\kappa^2-m^2) ,\nonumber\\ 
&&Q=\tilde p^2 -2i p_4 \kappa,\quad 
  \tilde p^2 \equiv p_4^2+v_F^2\p^2, \quad 
  p_4=2 \pi n T,\quad
   \p=|\bp|. \nonumber 
\end{eqnarray}
For the vacuum part, $\Pi^{\rm (vac)}$, one can directly use \Ref{Pmn} for Euclidean momenta, giving 
\begin{equation}
	\Pi^{\rm (vac)}_{00}=\frac{\alpha \p^2 \Phi(ip_4,\bp) }{\tilde p^2} \,,\qquad
	\Pi^{\rm (vac)}_{\rm tr}=\frac{\alpha  (p^2+\tilde p^2) \Phi(ip_4,\bp)}{\tilde p^2}
\,.\label{Pivac}
\end{equation}
The important feature of the representation \Ref{decomp1} is that it permits for an analytic continuation to the real (optical) frequencies from the Matsubara ones, and thus can be applied for investigation of the optical properties, surface plasmons and other effects in graphene at finite temperature and chemical potential.

The polarization tensor and, more generally, Feynman diagrams involving
$2+1$ dimensional fermions were considered in a number of papers.
The function $\Phi$ (for $v_F=1$) was first calculated in Ref.\ \cite{Appelquist}, while the pseudotensor part was discussed about the same time in the context of the parity anomaly\cite{Sem,Red}.
In this century, extensive calculations were done by the Kiev group and collaborators\cite{Gusynin1,Gusynin2,Gusynin3,Pyatkovskiy}, and results were rederived and analyzed in works by Klimchitskaya and collaborators, e.g. \cite{Klim2015a,Klim2015b,Klim2016}. The decomposition \Ref{decomp1} is a well known feature of polarization tensor in different theories, see e.g. \cite{shuryak}, and was applied for graphene at $\mu=0$, $T\ne0$ in \cite{Klim2015a} and $T=0$, $\mu\ne0$ in \cite{Bordag:2015gla}, and $T\ne0$, $\mu\ne0$ in \cite{BFV2015}.
The formulas (\ref{Pmn}), (\ref{Psi}) and (\ref{phi}) are consistent with those calculations.

\section{Physical effects in nanoribbons}\label{sec-phy}
We proceed with considering some applications of QFT in graphene nanodevices, namely graphene nanoribbons.

\subsection{Mean charge density}\label{sec-ch}
The calculation of the induced density of charge carriers, $n(x) = \langle j^0 (x)\rangle$, probably is the simplest application of the field theoretical methods for description of graphene, of course when $n(x)$ is non--trivial. It is the case for the nanoribbons. For the Berry-Mandragon ones, Eqs. \Ref{S} and \Ref{modes} can be used to obtain $n(x)$ from (\ref{j01}). 

Upon averaging over the cross section of the nanoribbon, it gives
\begin{equation}
   n=-\frac{i }{(2\pi)^2W} \sum_{\alpha,p_1}\int dp_0\, dp_2\,
        \frac 1{\mathcal{E}(\alpha,p_0,p_1,p_2)} \,,\label{j04}
\end{equation}
here $\mathcal{E}$ is given by \Ref{E}. This expression is UV divergent and calls for application of renormalization procedure, so uncommon for quantum mechanical treatment of graphene. In \cite{BFSV2014} the Pauli--Villars prescription was used. The physical motivation for the Pauli--Villars subtraction scheme is that
for a very large mass gap all fluctuations are frozen and do not contribute to quantum effects, as the mean charge density and conductivity, for example.

On this way it was obtained \cite{BFSV2014}
\begin{equation}
    n=\frac{  {\rm sgn}\, \mu}{\pi W v_F} \sum_{p_1>0}
    \sqrt{\mu^2-v_F^2\tk_0^2}\,\Theta(\mu^2-v_F^2\tk_0^2)\,,
\label{j07}
\end{equation}
where $\tk_0\equiv\tk_0(m)=\sqrt{p_1^2(m)+m^2}$. This expression is actually insensitive to the particular form of the dependence of $p_1=p_1^{(n)}(m)$ neither on index $n$ nor on the mass. Therefore, Eq.\ (\ref{j07}) is true for rather general boundary conditions, as long as there is no gapless mode, see more details in \cite{BFSV2014}.

It is also straightforward to calculate the quantum capacitance, $C_Q=e \partial Q/\partial \mu$, where $Q=e n$ is the induced charge density in the ribbon.
%\begin{equation}
 %  C_Q=\frac{e^2}{\pi W v_F} \sum_{k_1>0}
 % 	\frac{|\mu|}{\sqrt{\mu^2-v_F^2\tk_0^2}} \Theta(\mu^2-v_F^2\tk_0^2)\, .
%\label{C_Q}
%\end{equation}
A detailed analysis of the capacitance and its dependence on the gate/chemical potential is given in, e.g. \cite{Fang07,Shylau09},
whose results are in agreement with (\ref{j07}) though obtained via completely different approaches.

For graphene in a half-space subject to general boundary conditions the induced charge was considered in \cite{Nalimov} using a similar approach.

\subsection{Conductivity quantization in nanoribbons}\label{sec-DC}
With help of the fermion propagator \Ref{S}, and using \Ref{siPi} and \Ref{Pi_g}, the longitudinal optical conductivity $\sigma_{22}$ of a  Berry-Mondragon nanoribbon becomes 
\cite{BFSV2014}
\be
\S(\om)= \frac{2 e^2 v_F^2}{(2 \pi)^2W \omega} \int dp_0dp_2
    \sum_{p_1} \frac{\zeta(p_0)\zeta(p_0-\om)+v_F^2(p_2^2-p_1^2-m^2)}{\(\zeta^2(p_0)-v_F^2\varkappa^2\) \(\zeta^2(p_0-\om) -v_F^2\varkappa^2\)}\,,
    \label{S_xx}
\ee
A simple analysis shows that this expression is UV divergent. It can be handled with the Pauli-Villars regularization scheme.

\begin{figure}
\centering
\includegraphics[width=7cm]{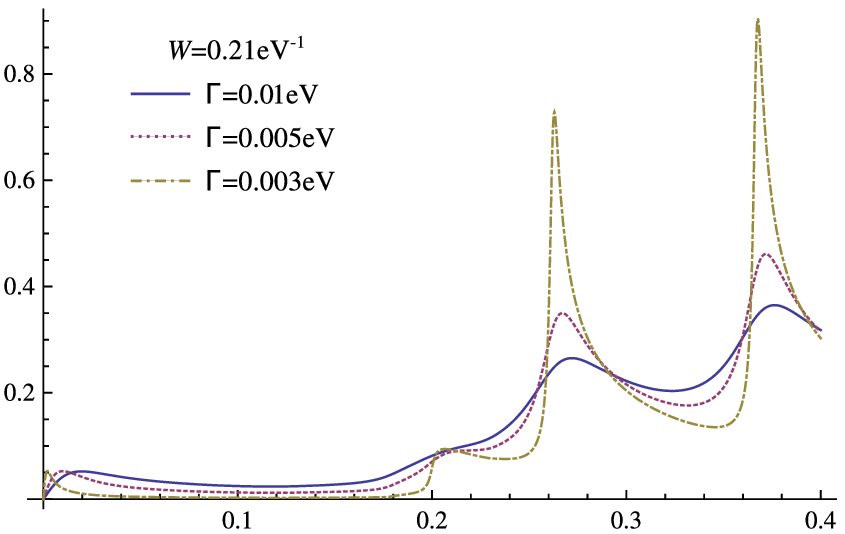}\includegraphics[width=7cm]{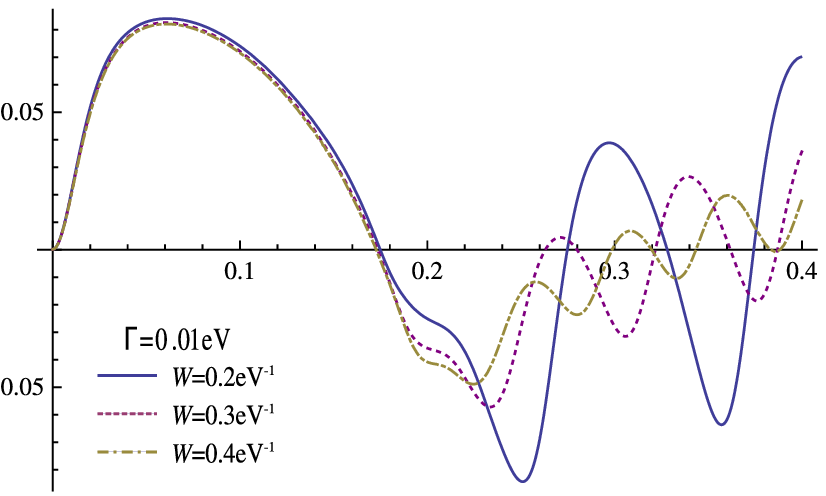}
\caption{
${\rm Re}\,\S^R $ (left) and ${\rm Im}\,\S^R $ (right) as functions of the frequency $\omega$[eV] at fixed chemical potential $\mu=0.1$eV, for different values of $\G$ and $W$, and for $m=0$, in units of $2 e^2/h$.
Reproduced from \cite{BFSV2014} with kind permission of The European Physical Journal (EPJ). 
}\label{ACcond}
\end{figure}

The result of this procedure is renormalized conductivity which is a cumbersome quadrature depending of the parameters of the model given explicitly in \cite{BFSV2014}. Its behavior is exemplified in Fig. \ref{ACcond} where the real and imaginary parts of the renormalized conductivity are presented as functions of the frequency at fixed chemical potential. One notes that optical absorption in the nanoribbons is comparatively small for frequencies smaller than $2\mu$, and shows clear resonance lines for higher values of the frequency. The peaks  in real and imaginary parts of conductivity correspond to those of the poles in the complex frequency plane lying at $\omega = 2 v_F p_1^{(n)}-(1+2i)\G$, for more details see \cite{BFSV2014}.

\begin{figure}
\centering \includegraphics[width=8cm]{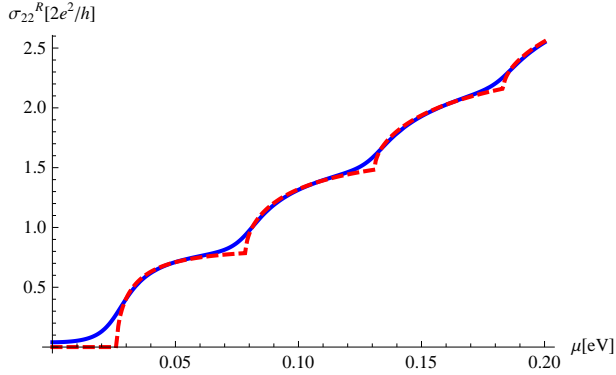}
\caption{Longitudinal DC conductivity of a Berry-Mondragon nanoribbion for a single fermion specie as a function of chemical potential $\mu$[eV], for $W=0.2$eV$^{-1}\simeq 40$nm, $m=0$,  $\G=10$meV. The exact expression (\ref{S_exact}) is given in solid--blue line,  while the dashed--red line corresponds to approximate result \Ref{S_app}, both in units of $2 e^2/h$. Reproduced from \cite{BFSV2014} with kind permission of The European Physical Journal (EPJ). }
\label{DCcond}
\end{figure}

The DC conductivity, describing a (linear) response of nanoribbon to a constant electric field, can also be obtained
\begin{eqnarray}
 \S^{R}(\om=0)&=&\frac{e^2 v_F}{2 \pi W  \G\mu}  \sum_{n}
        {\rm Im}\(\sqrt{v_F^2\tk_0^2(n)+(\G+i\mu)^2}- \frac{\G(\G+i\mu)}{\sqrt{v_F^2\tk_0^2(n)+(\G+i\mu)^2}}\)
	  \label{S_exact}\\
 &\underset{\G\to0}{\seq} & \frac{e^2 v_F}{2 \pi W  \G|\mu|}\sum_{p_1>0}
         \Theta(\mu^2-v_F^2\tk_0^2)\sqrt{\mu^2-v_F^2\tk_0^2}+O\(1/\sqrt\G\)\,.
    \label{S_app}
\end{eqnarray}
here $\tk_0 (n)=\sqrt{k_1^2(n)+m^2}$, and the second line corresponds to the limit of weak impurities, $\G\to0$. A remarkable property of the DC conductivity  \Ref{S_exact} is its quantization in a square-root manner, not  a step-wise one obtained for the conductance quantization in Landauer approach \cite{TitovPRL06}. Such behavior is in accordance with numerical simulations \cite{Mrneca2015}, and is induced by averaging the Fabry-Perot oscillations of the propagating modes between the leads, and not the presence of impurities. This behavior is given at the Fig. \ref{DCcond}.

The calculation of mobility and the minimal conductivity for Berry-Mondragon graphene nanorribbons follows immediately from \Ref{S_app}.

\section{Physical effects in infinite graphene}\label{sec-phy}
We turn now to investigate the QFT effects in infinite graphene layers.
Most of the interesting physics comes here from the connection between the polarization tensor $\Pi^{mn}$ and the in-plane conductivity of the layer, discussed in Sect. \ref{sec-pol}.

\subsection{Optical properties}
\label{sec-optical}
The first investigations of optical properties of graphene with QFT were based on the consideration of the linearly polarized electrodynamic potential \cite{Fialkovsky:2009wm}, \cite{FV}. The matching conditions \Ref{match} are readily solved in this case showing that a linearly polarized light will suffer both polarization rotation $\theta$ and intensity $\mathcal{I}$ reduction under passing a graphene layer \cite{Fialkovsky:2009wm}.
Indeed, for  a general  matter-free polarization operator \Ref{Pmn} one obtains 
\begin{equation}
    \theta= -\frac{{\rm Re}\sigma_{xy}}2+O(\alpha^2), \qquad
        \mathcal{I}= 1 - { {\rm Re}\sigma_{xx}}+O(\alpha^2)\,,
    \label{Tth_a}
\end{equation}
where we used (\ref{siPi}) to express the $\Pi^{mn}$ components through diagonal and Hall conductivities of graphene, and expanded the result in powers of the finite structure constant $\alpha$. 

For pristine graphene in absence of external fields the parity odd components of polarization operator cancel out and thus $\sigma_{xy}=0$. As shown in \cite{Fialkovsky:2009wm} in this case ${ {\rm Re}\sigma_{xx}}\simeq \alpha\pi$, and \Ref{Tth_a} thus gives a QFT confirmation of the predictions of Refs.\cite{ando,Fal,stauber,abs2} of universal absorption rate of $\alpha\pi \simeq 2.3$\%, that was confirmed of course by the famous experiment\cite{abs1}.

On the other hand, in presence of constant magnetic field, the P-odd term in $\Pi^{\mu\nu}$ will persist and the rotation of polarization of light is also possible, thus producing the Faraday effect. The polarization operator under such conditions was calculated in \cite{FV} showing a good agreement between the predictions of Dirac model for polarization rotation and experimental study \cite{Fara}. The latter revealed a  ``giant'' Faraday rotation angle of about $0.1$rad peaked at low frequencies for a magnetic field of about $7$~Tesla. Besides, in \cite{FV} there were other effects predicted, such as step-function like behavior of the rotation angle and the peaks at higher frequencies.

While for optical experiments it is quite natural to treat light in the basis of linear polarizations, for the future applications to the investigation of Casimir effect, it is desirable to consider also a different set of modes, namely TE and TM ones.%
\footnote{Though any electromagnetic field in vacuum can always be decomposed into a linear combination of two independent modes, their interaction with matter, graphene for instance, not always can be described independently from each other, see for instance \cite{MacDonald}.} Its advantage is that in the certain limits considered in Sec. \ref{sec-Cas} the interaction of graphene with TM mode will mimic the ideal metal case, while the one with TE mode will disappear. 

TE mode with the frequency $\omega$ propagating in vacuum along the $x^3$-axis from $x^3=-\infty$ is given by
\begin{eqnarray}
 && {\bf E}  =(-p_2 {\bf e}_1+p_1 {\bf e}_2) \omega \Psi(x^3) \\
 && {\bf H}  = i(p_1 {\bf e}_1+p_2 {\bf e}_2)  \Psi'(x^3)
    +{\bf e}_3 (p_1^2+p_2^2)   \Psi(x^3) 
\end{eqnarray}
while TM mode is
\begin{eqnarray}
&&  {\bf E}  =i(p_1 {\bf e}_1+p_2 {\bf e}_2)   \Phi'(x^3)
    +{\bf e}_3 (p_1^2+p_2^2)  \Phi(x^3) \\
&&  {\bf H}  = (p_2 {\bf e}_1 - p_1 {\bf e}_2) \omega   \Phi(x^3) 
\end{eqnarray}
where ${\bf e}_1,\ {\bf e}_2, {\bf e}_3$ are unit vectors. An overall factor
of $\exp (i(x^0 \omega+x^1p_1+x^2p_2))$ has been omitted for brevity.
To define the scattering data in the TE and TM sectors we take the
potentials in the form \be
    \Psi(x^3) = \begin{cases}
            e^{ip_3x^3}+r_{\rm TE}e^{-ip_3x^3}, \quad x^3<0 \cr
            t_{\rm TE}e^{ip_3x^3}, \quad x^3>0
            \end{cases},\qquad
    \Phi(x^3) = \begin{cases}
            e^{ip_3x^3}+r_{{\rm TM}}e^{-ip_3x^3}, \quad x^3<0 \cr
            t_{\rm TM}e^{ip_3x^3}, \quad x^3>0
            \end{cases}
\ee

The reflection and transmission coefficients can be derived now by solving 
matching conditions \Ref{match}. For the first time they were obtained in \cite{Fialkovsky:2011pu} and rederived in numerous papers afterwards. In terms of the polarization operator components they read (in Minkowsky space)
\be 
r_{\rm TM}=\frac{p  \Pi_{00}}{p \Pi_{00}  + 2 i \bp^2},
\qquad r_{\rm TE}= - \frac{ p^2 \Pi_{00}+ \bp^2 \Pi_{\rm tr}}
            {p^2 \Pi_{00} +  \bp^2 (\Pi_{\rm tr} - 2 i p)}.
\label{rTETM-grPi}
\ee
here $p=+\sqrt{\om^2-\bp^2}$.

Using a similar procedure, one is able to calculate easily the reflection coefficients for multylayered system of graphene, or topological insulators, embedded in dielectrics and  metamaterials of varying properties, e.g. \cite{graCas-compar,LopesGrushin2014, Woods2011}.

\subsection{The Casimir effect}\label{sec-Cas}
The Casimir effect\cite{Bordag:2001qi} is one of the very few macroscopical manifestations of the quantum nature of the classical objects and fields. It is not accessible via pure Quantum Mechanical approaches, and has no classical electromagnetic analog. As particular example, we consider a suspended graphene sample separated by the distance $a$ from a parallel plane ideal conductor.  Using the above described methods, different parameters describing the graphene sample can be taken into consideration --- mass $m$, chemical potential $\mu$ and temperature $T$.

Investigation of the Casimir effect is based nowadays on the Lifshitz\cite{Lifshitz} formula, which relates the optical properties of two interacting bodies to the free energy of the system
\begin{equation}
    {\mathcal F}
    =T\sum_{n=-\infty}^\infty\int\frac{d^2{\bp}}{8\pi^2} \ln [(1-e^{-2p a}r_{\rm
 \rm TE}^{(1)}r_{\rm \rm TE}^{(2)})
        (1-e^{-2p  a}r_{\rm \rm TM}^{(1)}r_{\rm \rm TM}^{(2)})] \,,
        \label{EL}
\end{equation}
where $p=\sqrt{\omega_n^2+{\bp}^2}$, and $\omega_n=2\pi n T$ are the
Matsubara frequencies.
$r^{(1,2)}_{\rm TE,TM}$ are the reflection coefficients for the TE and TM modes at
each of the two surfaces -- graphene and ideal conductor, in our case. For the latter one 
we have $r_{\rm \rm TM}^{(2)}=1$, $r_{\rm TE}^{(2)}=-1$. The reflection coefficients for graphene are given in \Ref{rTETM-grPi}, encoding its properties via the components of the polarization operator $\Pi=\Pi(m,\mu,T)$. Varying its dependence on these parameters we are able to obtain different limiting cases and investigate the behavior of the Casimir energy in transient regimes. 

As compared to diagrammatic QFT approaches the Lifshitz formula has the advantage of taking both boundaries non-perturbatively into account. It was shown\cite{Bordag:2009fz} that in  the case of pristine graphene at zero temperature, $m=\mu=T=0$,  it coincides with the first non-trivial diagram describing the free energy for the system \Ref{Seff}. It was found that both approaches give consistent result of order $2.7\%$ of the Casimir interaction between two ideal metals.

Some unexpected features appear at non-zero temperature\cite{Fialkovsky:2011pu}, but still with $m=\mu=0$. The free energy (\ref{EL}) grows with increase of the dimensionless parameter $aT$ as compared to the interaction between two ideal metals, and, at the large $T$ asymptotic we have
\begin{equation}
{\mathcal{F}}  \seq_{T\to \infty} \frac12 {\mathcal F}_{\rm id}
	\equiv -\frac{T\zeta(3)}{16\pi a^2} \,,
\label{Fas}
\end{equation}
which is just a half of the interaction between two ideal metals in the same regime, or is the same value as for non-ideal metals described by the Drude model! Thus, the Casimir interaction of graphene at high temperature is extremely strong. This agrees qualitatively with Ref.\ \cite{Gomez} where the Casimir interaction of two graphene samples
was considered.

Similarly surprising is the Casimir energy for doped graphene samples, i.e. for large $|\mu|$. Supposing that $a$ is large enough to neglect all terms with non-zero Matsubara frequencies $n\ne 0$ in (\ref{EL}), it was derived  \cite{BFV2015} that the Casimir interaction in the limit $\mu\to \infty$ reaches the very same value of $1/2$ of the ideal metal - ideal metal one,
\be
   {\mathcal{F}} \seq_{\mu\to\infty}
         \frac12 {\mathcal F}_{\rm id}.
	\label{1halve}
\ee

Both limits \Ref{Fas} and \Ref{1halve} are invoked by the very specific structure of the reflection coefficients \Ref{rTETM-grPi} that guarantees that only the TM mode contributes to the Casimir interaction in these limits. Indeed, in both of them, the value of $r_{\rm TM}$ tends to $1$ of an ideal conductor, while $r_{\rm TE}$ becomes of order $O(\alpha)$.  Eqs.  \Ref{Fas}, \Ref{1halve} give a very rough idea on how far the enhancement of the Casimir effect with $\mu$ or $T$ might go. While high temperatures are accessible in experiments though cumbersome to deal with, consideration of $|\mu|$ exceeding a couple of eV in the framework of the Dirac model it hardly makes much sense.

Numerical analysis at $T=300$K and $m=0$ show \cite{BFV2015} that at distances about $100$--$300$ nanometers the Casimir effect between a perfect metal plate and doped graphene is highly enhanced even for relatively moderate values of the chemical potential. In particular, for $\mu=0.8$eV the force between a doped graphene layer and an ideal metal is almost $60$\% higher then that for a pristine one.  It was also noticed that  the effect is more pronounced the more derivatives one calculates of the energy. Thus, the ratio of the energy density at $\mu=0.8$eV to its pristine value is $1.52$ at maximum, for the force ${\cal P}=-\partial {\cal F}/\partial a $ it is $1.54$ and for the force gradient  $ {\cal G}=\partial {\cal P}/\partial a $  it reaches $1.56$. 

\begin{figure}
\centering \includegraphics[width=8cm]{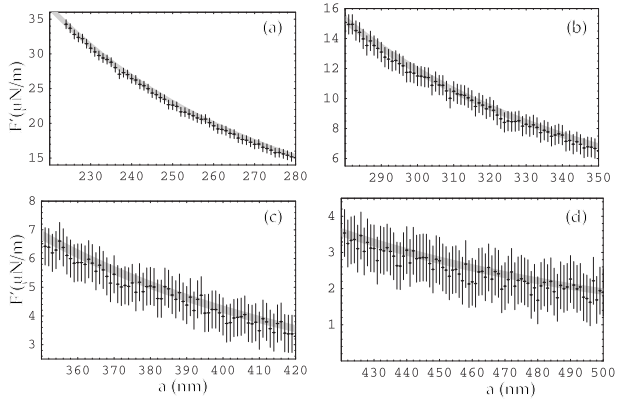}
\caption{The experimental data for
the gradient of the Casimir force
between an Au-coated sphere and graphene deposited on a
SiO${}_2$ film covering a Si plate are shown as crosses plotted at a 67\% confidence level
over different separation regions. The gray bands present the
theoretical force gradients computed using the exact
reflection coefficients for graphene on a
multilayered substrate. 
%\footnote{
Reprinted with permission from \cite{graCas-compar}. Copyright (2014) by the American Physical Society.
}
\label{graCas}
\end{figure}

The experimental status of Casimir interaction in graphene systems is as follows. 
Using a dynamic atomic force microscope the gradient of the Casimir force between an Au-coated sphere and a graphene sheet deposited on a SiO$_2$ film covering a Si plate  was measured \cite{graCas-exp}. A comparison of the measurements with the predictions of QFT encoding graphene properties into the Casimir interaction via the polarization operator \Ref{Pi_g},\Ref{decomp1} was performed in \cite{graCas-compar}. Unlike \Ref{rTETM-grPi} obtained for a freestanding graphene in the latter work the reflection coefficients of a multi-layered systems were obtained, and the gradient of the Casimir force was calculated using the Lifshitz formula with those reflection coefficients. It was showed in \cite{graCas-compar} that the theoretical results are in very good agreement with the experimental data. On Fig. \ref{graCas} the comparison of theoretical predictions and the experimental results are given. However, tempretature and chemical potential dependences of the Casimir interaction were not resolved.

We conclude this Section with some references. Other papers which study
the Casimir effect for graphene are \cite{Woods2011,Dobson,Sernelius,Ali,Woods2012}. A number of papers has been dedicated to the investigation of Casimir effect in related systems: for for topological insulators \cite{LopesGrushin2014,GrushinCorjito2011} and graphene subject to magnetic field \cite{MacDonald} and under the strain \cite{Phan2014}. In these cases, the lack of symmetry in the system does not permit an independent propagation of two separate modes (like TE and TM in \Ref{rTETM-grPi}) and one has to consider a $2 \times 2$ reflection matrix, see e.g. \cite{RosaDalvit2008}.

More details on the presents status of Casimir effect in graphene can be found
in Ref.\ \cite{Mar}.

\section{Conclusions}
The aim of this paper, that may be viewed as an updated version of our previous review \cite{FV2011} on the same subject,  is to show the development of the quantum field theory calculations based on the Dirac model of quasiparticles in describing the physics of graphene.  The field theoretical methods show themselves readily applicable and reliable in this field mainly due to the well known fact that the Dirac model at small momenta gives a very good approximation for the tight binding model.

Finally, we have to mention at least some of the subjects left aside of this review. One of them, is the rigorous mathematical formulation of the physical boundary conditions being imposed on the graphene nanoribbons and other graphene devices (nano dots, etc.) Some types of such conditions, such as the Bery-Mondragon ones above, are very well understood both in physical and mathematical sense of the word, and even the families of appropriate conditions are investigated, e.g. \cite{BSrib,Igor2009}. Still, the compatibility of others, such as Zig-Zag ones\cite{Akhmerov08}, with quantum field theory is yet an open question, due to presence of infinite number of zero modes. 

The other appealing area is the application of quantum field theory in curved spaces to the better treatment of graphene both free standing and the strained one. While in the former case the natural ripples of graphene layers are responsible for the effective curvature of the space, it is the non-uniform strain that modifies the Dirac action in the latter. 

All alone stay the applications of field theoretical methods to topologically-nontrivial phases of matter such as Weyl semimetals and Topological Insulators. This field contains quite a number of open interesting problems due to its relative novelty. 

Of course, this review is far from being a comprehensive study of all possible applications of the field-theoretical methods in graphene physics, providing only the general overview of some of the QFT applications.

\section*{Acknowledgments}
We are grateful to M.~Bordag, G.~Beneventano, M.~Santangelo and V.~Marachevsky for collaboration and for fruitful discussions, and to the Organizers
of QUARKS 2016 for making this enjoyable event. This work was supported
in parts by FAPESP (D.V.V.) and by CNPq (D.V.V.).

\end{document}